\documentclass[aps,twocolumn,prb]{revtex4}

\usepackage{graphicx}
\usepackage{amssymb}
\usepackage{amsmath}

\begin{document}

\title{Accurate non-covalent interaction energies via an efficient MP2 scaling procedure}
\author{E. Fabiano}
\affiliation{Istituto Nanoscienze-CNR, Euromediterranean Center for Nanomaterial Modelling and Technology (ECMT), via Arnesano 73100, Lecce}
\affiliation{Center for Biomolecular Nanotechnologies @UNILE, Istituto Italiano di Tecnologia, Via Barsanti, I-73010 Arnesano, Italy}
\author{F. Della Sala}
\affiliation{Istituto Nanoscienze-CNR, Euromediterranean Center for Nanomaterial Modelling and Technology (ECMT), via Arnesano 73100, Lecce}
\affiliation{Center for Biomolecular Nanotechnologies @UNILE, Istituto Italiano di Tecnologia, Via Barsanti, I-73010 Arnesano, Italy}
\author{I. Grabowski}
\affiliation{Institute of Physics, Faculty of Physics, Astronomy and Informatics, Nicolaus Copernicus University, Grudziadzka 5, 87-100 Torun, Poland}

\begin{abstract}
Using the observed proportionality of CCSD(T) and MP2 correlation interaction
energies [I. Grabowski, E. Fabiano, F. Della Sala, Phys. Chem. Chem. Phys. 15,
  15485 (2013)] we propose a simple scaling procedure to compute accurate
interaction energies of non-covalent complexes. Our method makes use of MP2
and CCSD(T) correlation energies, computed in relatively small basis sets, 
and fitted scaling coefficients
to yield interaction energies of almost complete basis set limit 
CCSD(T) quality. Thanks to the good
transferability of the scaling coefficients involved in the calculations,
good results can be easily obtained for different intermolecular 
distances.  
\end{abstract}

\maketitle

\section{Introduction}
Non-covalent interactions play a fundamental role in chemistry and 
chemical-physics because of their importance in many 
phenomena including, among others, biochemistry, solvation, supermolecular organization,
and molecular recognition. For this reason they have been subject of intense research
\cite{noncov_book,super_book,hohenstein12,hobza12}.
However, because the strength of non-covalent interactions 
is one or two order of magnitude smaller than that of typical 
covalent interactions, ranging from few tens 
to few tenths of kcal/mol, special care is required for 
accurate quantitative studies.

From a computational point of view this implies the need for 
high-level correlated wave function methods, being capable to describe
correlation effects with good accuracy \cite{hohenstein12,hobza12}. 
Additional computational issues, such as the basis set 
superposition and the basis set incompleteness errors,
must be properly taken into account as well
\cite{tao01,bsse_rev,cbs_chapter,feller11,rpa_extra,scheiner12,richard13}.
Finally, it must be considered that, because of the rather weak nature of 
non-covalent bonds, the description of non-covalent complexes cannot
be restricted to the equilibrium geometry, but, in many cases, 
must be extended to include a more or less extended portion
of the potential energy surface (PES). 

Actually the method of choice for an accurate computational description 
of non-covalent complexes is the coupled cluster single and double with 
perturbative triple (CCSD(T)) approach \cite{purvis82,scuseria88,pople87,ccsdt}
with a complete basis set (CBS) description \cite{cbs_chapter}.
This method provides in fact interaction energies with subchemical accuracy and yields
errors often below 1 kcal/mol \cite{hobza12}.
Nevertheless, the CCSD(T) method is computationally very demanding,
scaling as $\mathcal{O}(N^7)$, thus its applicability is limited to small complexes.
On the other hand, lower level approaches, such as M\o ller-Plesset second-order
perturbation theory (MP2) \cite{mp2} or even CCSD(T) calculations with a relatively small 
basis set, fail to describe various types of non-covalent interactions with the 
necessary accuracy.
Therefore, continuous effort is dedicated to the development of new methods able of
providing reliable results for non-covalent complexes at a reduced computational cost.

Recently one step in this direction has been performed by proposing a simple
non-empirical scaling procedure for MP2 interaction energies which allows to
obtain a full dissociation curve of CCSD(T) quality at the cost of a single
CCSD(T) calculation \cite{cos_scal}. This method is based on an observed
proportionality of the MP2 and CCSD(T) interaction energies which is
maintained, with almost a constant factor, over a wide range on inter-molecular distances.
Thus, accurate dissociation curves can be computed, with a precision of few
hundreds of mHa, for many 
non-covalent complexes with a reduced effort.
However, for high accuracy the proposed scaling procedure still requires one 
CCSD(T)/CBS calculation, which may be out of reach for larger scale 
applications.

In this paper we address this latest issue and propose a new methodology which
joins the above mentioned scaling procedure with a basis set extrapolation scheme,
so that finally highly accurate interaction energies can be computed
by the sole requirement of relatively low-level calculations (e.g. MP2 and/or
CCSD(T) with moderate basis sets). In this way a practical and efficient
tool for large-scale application can be obtained.

\section{Method}
The main quantity of interest in this work
is the correlation interaction energy.
For calculations carried out with method $X$
and using the basis set aug-cc-pV$n$Z it is denoted 
$\mathcal{E}_X^{(n)}$ and defined as the difference
of the correlation energies ($E_{c,X}^{(n)}$, calculated using method $X$ 
and basis set aug-cc-pV$n$Z, with $n=2,3,4,5$ i.e. D,T,Q,5) for a complex $AB$ 
and its constituent fragments $A$ and $B$, i.e.
\begin{equation}
\mathcal{E}_X^{(n)} = E_{c,X}^{(n)}[AB] - E_{c,X}^{(n)}[A] - E_{c,X}^{(n)}[B]\ .
\end{equation}

According to Ref. \cite{cos_scal} the CCSD(T) 
correlation interaction energy of a
non-covalent complex at any inter-molecular 
separation $R$ is related to the
MP2 correlation interaction energy by the simple equation
\begin{equation}\label{e1}
\mathcal{E}_\mathrm{CCSD(T)}^{(n)}(R) \approx c^{(n)}\mathcal{E}_\mathrm{MP2}^{(n)}(R)\ ,
\end{equation}
where the factor $c^{(n)}$ is evaluated as
\begin{equation}\label{e2}
c^{(n)} = \frac{\mathcal{E}_\mathrm{CCSD(T)}^{(n)}(\tilde{R})}{\mathcal{E}_\mathrm{MP2}^{(n)}(\tilde{R})}\ ,
\end{equation}
with $\tilde{R}$ being any reference point (for example, but not
necessarily, the equilibrium inter-molecular separation $R_0$).
Equations (\ref{e1}) and (\ref{e2}) allow to
compute a full PES (with $N$ points) of CCSD(T) quality at the cost of a single
CCSD(T) calculation (at distance $\tilde{R}$) and several (i.e. $N$)
MP2 calculations.

\begin{figure}
\includegraphics[width=\columnwidth]{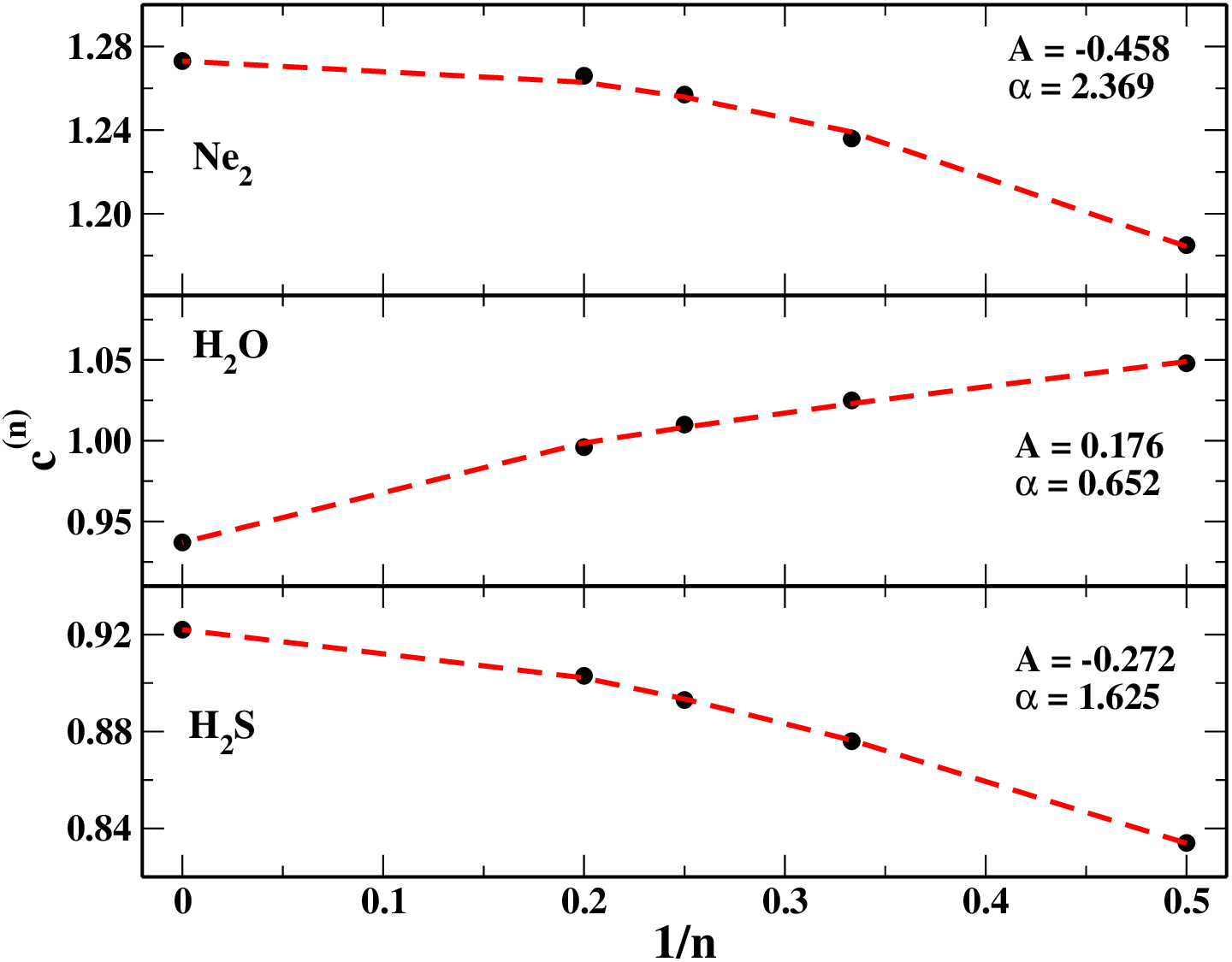}
\caption{\label{fig1} Scaling factors $c^{(n)}$ as functions 
of $1/n$ (where $n$ defines the basis sets aug-cc-pV$n$Z)
for some non-covalent complexes.
The dashed line shows the fit according to Eq. (\ref{e3}).}
\end{figure}

When the CBS limit is considered, which formally corresponds to $n=\infty$,
the scaling factor of Eq. (\ref{e2}) shall be computed as
$c^{(\infty)}=\mathcal{E}_\mathrm{CCSD(T)}^{(\infty)}/\mathcal{E}_\mathrm{MP2}^{(\infty)}$,
where both $\mathcal{E}_\mathrm{CCSD(T)}^{(\infty)}$ and $\mathcal{E}_\mathrm{MP2}^{(\infty)}$
have to be obtained via appropriate extrapolation formulas
(e.g. Refs. \cite{truhlar1,truhlar2}). However, the
bottleneck of this computational approach is certainly the calculation
of $\mathcal{E}_\mathrm{CCSD(T)}^{(\infty)}$. To avoid this we consider
an alternative way to compute $c^{(\infty)}$. To this end we assume that
the basis set evolution of the proportionality factor $c^{(n)}$ is described by the
ansatz
\begin{equation}\label{e3}
c^{(n)} = c^{(\infty)} + An^{-\alpha}\ ,
\end{equation}
where $A$ and $\alpha$ are assumed to be two system-dependent constants
(note that they do not depend on $n$). A rationalization for this
ansatz is given in \ref{appa}. A graphical impression of its
accuracy for some typical non-covalent complexes is presented in Fig. \ref{fig1}.
This figure shows in fact, for some selected complexes, the evolution
of the values of the scaling factors $c^{(n)}$ with the
basis set as well as the fit obtained via Eq. (\ref{e3}).
We see that the postulated power behavior reproduces very
well the trend of the computed scaling factors from
$1/n=0.5$ (aug-cc-pVDZ) to $1/n=0$ (CBS limit). 

In practice, we apply Eq. (\ref{e3}) for two given 
values of $n$ (we use $n=2$ and $n=3$) so that 
the CBS proportionality factor is given by the simple two-point formula
\begin{equation}\label{e4}
c^{(\infty)}_{approx} = \frac{c^{(3)}3^\alpha - c^{(2)}2^\alpha}{3^\alpha - 2^\alpha}\ ,
\end{equation}
where the only unknown parameter, so far, is $\alpha$.
Equation (\ref{e4}) allows to obtain an estimate for 
$c^{(\infty)}$ avoiding the
computationally expensive calculation of $\mathcal{E}_\mathrm{CCSD(T)}^{(\infty)}$, 
requiring at most the evaluation of $\mathcal{E}_\mathrm{CCSD(T)}^{(3)}$, which is
necessary to obtain $c^{(3)}$ from Eq. (\ref{e2}). 
On the other hand, the value of $\alpha$ should be, in principle
system dependent. However, following Ref. \cite{truhlar2},
we will fix it by fitting to some appropriate training set
(see later on). Hence, Eq. (\ref{e4}) becomes a universal
formula for obtaining $c^{(\infty)}_{approx}$.

Using Eq. (\ref{e4}) it is possible to obtain a good approximation
of $\mathcal{E}^{(\infty)}_\mathrm{CCSD(T)}$
in a very efficient way via Eq. (\ref{e1}). 
Nevertheless, two possible variants can be considered:
\begin{itemize}
\item In a first approach, that we denote M1, we will use the best
estimate available for $\mathcal{E}_\mathrm{MP2}^{(\infty)}$,
in practice, a cubic extrapolation formula \cite{rpa_extra,halkier98} 
using aug-cc-pVQZ and aug-cc-pV5Z data (other extrapolation formulas will not
be considered here, since they will yield only slightly different results).
Hence, the M1 correlation interaction energy is computed as
\begin{equation}
\mathcal{E}_{M1} = c^{(\infty)}_\mathrm{M1}\frac{\mathcal{E}^{(5)}_\mathrm{MP2}5^{3}-\mathcal{E}^{(4)}_\mathrm{MP2}4^3}{5^3 -4^3}\ ,
\end{equation}
where $c^{(\infty)}_{M1}$ is given by Eq. (\ref{e4}) with 
$\alpha$ fitted on a training set to reproduce reference 
values of the $c$ coefficient at the reference CBS limit.
Within this approach, to compute a full PES of $N$ points with CCSD(T)/CBS quality,
the most computationally intensive steps required are one
CCSD(T)/aug-cc-pVTZ calculation and $N$
MP2 calculations with large basis set (aug-cc-pV5Z in our case),
one for each point of the PES.
\item A cheaper method, which will be denoted M2, 
can be obtained using  a simpler formula to evaluate $\mathcal{E}_\mathrm{MP2}^\infty$.
Doing this by means of the approach of Ref. \cite{truhlar2}
we obtain
\begin{equation}\label{m2_eq}
\mathcal{E}_\mathrm{M2} = c^{(\infty)}_\mathrm{M2}\frac{\mathcal{E}^{(3)}_\mathrm{MP2}3^{\gamma}-\mathcal{E}^{(2)}_\mathrm{MP2}2^\gamma}{3^\gamma -2^\gamma}\ ,
\end{equation}
where the value of the coefficient $\gamma=1.91$ is taken from Ref. \cite{truhlar2}
and $c^\infty_\mathrm{M2}$ is given again by Eq. (\ref{e4})
with a proper value of $\alpha$ (see below).
In this case only $\mathcal{E}_\mathrm{MP2}^{(3)}$ calculations instead of
$\mathcal{E}_\mathrm{MP2}^{(5)}$ are 
required at most. Thus, the global
computational cost of the whole procedure is practically reduced
to the calculation of one $\mathcal{E}_\mathrm{CCSD(T)}^{(3)}$
(to compute $c^{(3)}$ in Eq. (\ref{e4})).
Of course, in this case to obtain high accuracy it is not possible
to use the same $\alpha$ value as for the M1 method (hence,
we used the subscripts M1 and M2 to denote the scaling factors in
the two methods). On the contrary,
it must be determined by fitting to the CBS values of CCSD(T)
reference correlation interaction energies, and not 
CBS limit $c$ values, in order to effectively take into account
also the possible inaccuracies on the extrapolated MP2 energy. 
\end{itemize}
\begin{table}
\begin{center}
\caption{\label{tab1}Computed and extrapolated values of the proportionality factors $c^{(n)}$ for the systems of the training set.}
\begin{ruledtabular}
\begin{tabular}{lrrrrr}
System & $c^{(2)}$ & $c^{(3)}$ & $c^{(\infty)}$ & $c^{(\infty)}_\mathrm{M1}$ & $c^{(\infty)}_\mathrm{M2}$ \\
\hline
He$_2$ & 1.262 & 1.261 & 1.248 & 1.261 & 1.261 \\
Ne$_2$ & 1.185 & 1.236 & 1.273 & 1.274 & 1.253 \\
He-Ne   & 1.229 & 1.266 & 1.274 & 1.294 & 1.279 \\ 
Ar$_2$ & 0.860 & 0.905 & 0.913 & 0.939 & 0.921 \\
(H$_2$O)$_2$ & 1.048 & 1.031 & 0.937 & 1.019 & 1.026 \\
(H$_2$S)$_2$ & 0.834 & 0.876 & 0.922 & 0.907 & 0.890 \\
(HF)$_2$ & 1.184 & 1.088 & 0.924 & 1.017 & 1.056 \\
(HCl)$_2$ & 0.791 & 0.846 & 0.853 & 0.887 & 0.865 \\
\end{tabular}
\end{ruledtabular}
\end{center}
\end{table}
\begin{table}
\begin{center}
\caption{\label{tab2}Extrapolated correlation interaction 
energies (kcal/mol) for the systems of the training set computed with different
methods. Reference values are CCSD(T)/CBS results. The last line reports the mean 
absolute error (MAE) and the mean absolute relative error (MARE).}
\begin{ruledtabular}
\begin{tabular}{lrrrrr}
System & MP2-23 & CCSD(T)-23 & M1 & M2 & Ref \\
\hline
He$_2$ & -0.03 & -0.04 & -0.04 & -0.04 & -0.04 \\ 
Ne$_2$ & -0.09 & -0.11 & -0.11 & -0.11 & -0.11 \\ 
He-Ne & -0.06 & -0.08 & -0.08 & -0.08 & -0.08 \\ 
Ar$_2$ & -0.67 & -0.61 & -0.63 & -0.61 & -0.61 \\ 
(H$_2$O)$_2$ & -1.36 & -1.42 & -1.39 & -1.38 & -1.27 \\ 
(H$_2$S)$_2$ & -2.21 & -2.04 & -2.00 & -2.04 & -2.04 \\ 
(HF)$_2$ & -0.92 & -0.86 & -0.93 & -0.87 & -0.85 \\ 
(HCl)$_2$ & -2.22 & -2.01 & -1.97 & -2.01 & -1.90 \\ 
          &        &         &        &        & \\
MAE &  0.09 &  0.04 &  0.04 &  0.03 & \\
MARE      &  14.1\%& 3.7\% & 3.6\% & 3.4\% & \\
\end{tabular}
\end{ruledtabular}
\end{center}
\end{table}

\section{Computational details}
All calculations have been performed using the PSI4 code
\cite{psi4} and aug-cc-pVnZ basis sets \cite{basis1,basis2,basis3,basis4}
with $n=$ D, T, Q, and 5 (i.e. =2, 3, 4, and 5). 
In all cases frozen core and resolution of identity \cite{ri} (for both
Hartree-Fock and correlated calculations) approximations have been 
employed. All results include counterpoise correction \cite{cp}.
{In all calculations}, unless otherwise stated, the
extrapolation to the CBS limit has been performed using a
two-point cubic formula based on aug-cc-pVQZ and aug-cc-pV5Z data
\cite{rpa_extra,halkier98} (the corresponding results
are indicated with the label method/CBS. On the other hand,
extrapolated results denoted MP2-23 and CCSD(T)-23 have been obtained using the
method and parameters reported in Ref. \cite{truhlar2}.
A similar notation is used for the scaling coefficients $c^{(n)}$
(see Eq. (\ref{e2})).

For the parametrization and testing of the methods
defined in this work we considered various small
non-covalent complexes having different interaction characters.
These include He$_2$, Ne$_2$, He-Ne, Ar$_2$, Ne-Ar, CH$_4$-Ne
(dispersion interaction; DI), (H$_2$O)$_2$, (HF)$_2$, NH$_3$-H$_2$O,
(NH$_3$)$_2$ (hydrogen bond; HB), (H$_2$S)$_2$, (HCl)$_2$, H$_2$S-HCl
(dipole-dipole interaction; DD), HCN-ClF, NH$_3$-F$_2$ (charge-transfer
interaction; CT), and LiH-HF (dihydrogen interaction; DH).
The geometries of the complexes were taken from Refs.
\cite{cos_scal,laricchia13,zhao05,zhao05_2,wesolowski96,dihydrogen}.
Finally, in addition, we considered the systems from the
S22$\times$5 test set \cite{grafova10}.

As reference benchmark data CCSD(T)/CBS results
obtained extrapolating from aug-cc-pVQZ and aug-cc-pV5Z
calculations or accurate CCSD(T)/CBS results from literature (when indicated). 
These accurate energies are employed to assess the accuracy of
all the methods based on -23 extrapolation, i.e. our own methods as well 
as MP2-23 and CCSD(T)-23 calculations. Moreover, for completeness,
we have considered also
SCS(MI)-MP2/CBS \cite{scsmi} and MP2.5/CBS \cite{mp2.5} 
interaction energies. These
methods were in fact developed to ``reproduce'' CCSD(T)
results at a cost close to MP2 calculations. Thus, it 
is in the same spirit as the ones developed in the present work.

\subsection{Parametrization}
To perform the fit required to fix the parameter $\alpha$
in both the M1 and M2 methods, we consider
a training set composed of the following non-covalent complexes:
He$_2$, Ne$_2$, He-Ne, Ar$_2$ (dispersion complexes), 
(H$_2$O)$_2$, (HF)$_2$ (hydrogen bond complexes), and (H$_2$S)$_2$, (HCl)$_2$ 
(dipole-dipole complexes), at equilibrium geometry.
The values of $c^{(2)}$, $c^{(3)}$, and $c^{(\infty)}$ for the various systems
are reported in Table \ref{tab1}.
Note that in Table \ref{tab1} $c^{(\infty)}$ denotes the ``true''
CBS limit value of the scaling factor, computed as the ratio
of the extrapolated CCSD(T)/CBS and MP2/CBS energies.

Considering method M1, we need to fit Eq. (\ref{e4}) to the 
$c^{(\infty)}$ values. Thus, we minimized the target function
\begin{equation}
\sigma_\mathrm{M1}(\alpha) = \sum_i\left(c^{(\infty)}_\mathrm{M1}(\alpha)[i]-c^{(\infty)}[i]\right)^2\ ,
\end{equation}
where the sum runs over all the systems in the training set and
the notation $c^{(\infty)}_\mathrm{M1}(\alpha)[i]$ indicates that
we consider the $c^{(\infty)}_\mathrm{M1}$ value relative to system $i$ 
and computed at a given value of $\alpha$.
Doing so we obtained
$\alpha=2.1$ and the $c^{(\infty)}_\mathrm{M1}$ values reported in
Table \ref{tab1}. The fitted coefficients show a mean absolute error
(MAE) of 0.035 and a mean absolute relative error (MARE) of 3.7\%
with respect to the $c^{(\infty)}$ ones.
The corresponding correlation interaction energies, obtained using the
M1 approach, are listed in Table \ref{tab2}.
They display a good agreement with the reference CCSD(T)/CBS energies
having a MAE of 0.06 mHa and a MARE of 3.6\%, being similar
to CCSD(T)-23 extrapolated energies.

To fit for the M2 method we consider instead, as discussed before, 
as target values the reference CCSD(T)/CBS correlation interaction 
energies of Table \ref{tab2}, hence we minimize
\begin{equation}
\sigma_\mathrm{M2}(\alpha) = \sum_i\left(\mathcal{E}_\mathrm{M2}(\alpha)[i]-\mathcal{E}^{(\infty)}_\mathrm{CCSD(T)/CBS}[i]\right)^2\ ,
\end{equation}
where $\mathcal{E}_\mathrm{M2}(\alpha)[i]$ depends on $\alpha$ through
$c^{(\infty)}_\mathrm{M2}$.
This yields $\alpha=3.4$ and the 
scaling factors displayed in the last column of
Tab. \ref{tab1}. When used in the M2 approach
these yield the final M2 energies
reported in Tab. \ref{tab2}. These energies show by construction a
very good agreement with reference values, having a MAE of only 0.05 mHa.

\section{Results}
\begin{figure}
\includegraphics[width=\columnwidth]{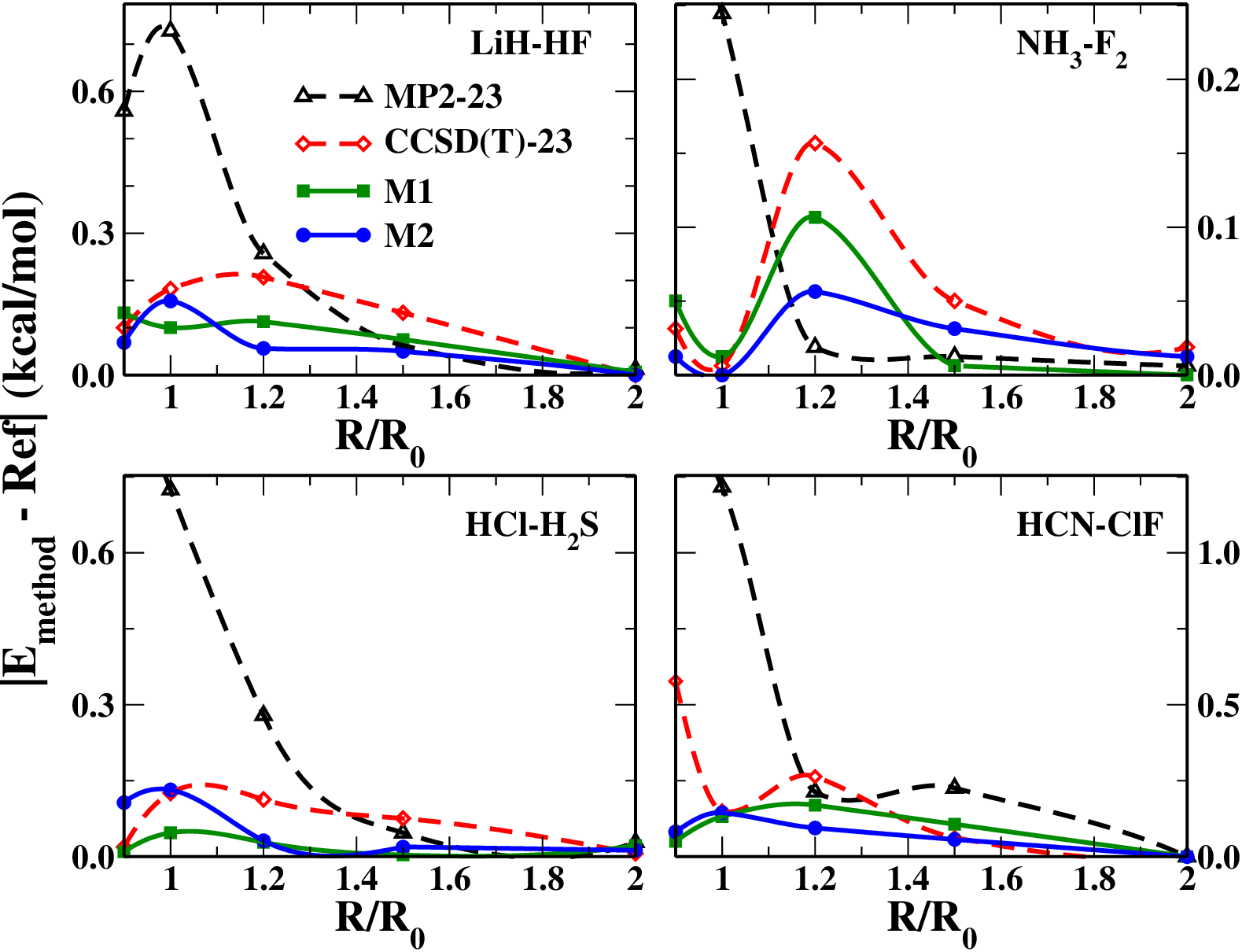}
\caption{\label{fig_pes}Absolute error with respect to reference values (CCSD(T)/CBS)
of correlation interaction energies (kcal/mol) computed with different methods and at different intermolecular distances $R$ ($R_0$ denotes the equilibrium distance) for some non-covalent complexes.}
\end{figure}

To test the methods developed in the previous sections
we applied them to the computation of the interaction energies of several
non-covalent complexes. For completeness we considered energies both
at the equilibrium distance $R_0$ and at elongated distances equal to 1.2$R_0$
and 1.5$R_0$.
The values of the correlation interaction energies, obtained from various methods, 
are reported in Table \ref{tab3}. In addition Fig. \ref{fig_pes} reports the
absolute error on the correlation interaction energy as obtained from different methods
for some complexes, taken as examples, including also shorter (i.e. 0.9$R_0$) and
larger (i.e. up to 2.0$R_0$) distances.
\begin{table*}
\begin{center}
\caption{\label{tab3}Computed and extrapolated correlation interaction energies (kcal/mol) for the systems of the test set at different intermolecular distances ($R_0$ denotes equilibrium distance). Reference values are CCSD(T)/CBS results. The label SCS(MI) stands for SCS(MI)-MP2/CBS. In the second column the interaction character for each complex is reported. The last line reports the mean absolute error (MAE) and the mean absolute relative error (MARE).}
\begin{ruledtabular}
\begin{tabular}{llrrrrrrr}
System & Interaction & MP2-23 & SCS(MI) & MP2.5/CBS & CCSD(T)-23 & M1 & M2 & Ref. \\
\hline
\multicolumn{9}{c}{$R=R_0$}\\
HCN-ClF & CT & -3.90 & -2.72 & -3.11 & -2.83 & -2.81 & -2.82 & -2.68 \\
NH$_3$-F$_2$ & CT & -2.33 & -1.56 & -2.08 & -2.08 & -2.10 & -2.09 & -2.09 \\
LiH-HF & DH & -3.60 & -2.65 & -2.85 & -3.06 & -2.97 & -3.03 & -2.87 \\
NH$_3$-H$_2$O & HB & -2.17 & -1.72 & -2.09 & -2.08 & -2.06 & -2.10 & -1.96 \\
(NH$_3$)$_2$ & HB & -1.70 & -1.27 & -1.64 & -1.66 & -1.60 & -1.61 & -1.57 \\
HCl-H$_2$S & DD &-3.27 & -2.45 & -2.97 & -2.67 & -2.59 & -2.68 & -2.55 \\
CH$_4$-Ne & DI &-0.28 & -0.17 & -0.29 & -0.30 & -0.33 & -0.29 & -0.29 \\
Ne-Ar & DI & -0.29 & -0.08 & -0.29 & -0.24 & -0.24 & -0.24 & -0.24 \\
        & &        &   &      &    &     &         & \\
MAE & &  0.41 &  0.21 &  0.14 & 0.09 &  0.06 &  0.08 & \\ 
MARE	&    & 18.8\% & 22.7\% & 7.9\% & 4.2\% & 4.0\% & 3.7\% & \\
\hline
\multicolumn{9}{c}{$R=1.2R_0$}\\
HCN-ClF & CT  & -1.12 & -1.40 & -0.84 & -0.64 & -1.07 & -0.81 & -0.90 \\
NH$_3$-F$_2$ & CT & -0.76 & -0.81 & -0.68 & -0.58 & -0.85 & -0.68 & -0.74 \\
LiH-HF & DH & -1.99 & -2.26 & -1.80 & -1.52 & -1.84 & -1.68 & -1.73 \\
NH$_3$-H$_2$O & HB & -1.02 & -1.10 & -0.98 & -0.91 & -1.09 & -0.99 & -0.93 \\
(NH$_3$)$_2$ & HB & -0.78 & -0.78 & -0.76 & -0.68 & -0.83 & -0.77 & -0.70 \\
HCl-H$_2$S & DD & -1.36 & -1.39 & -1.22 & -0.97 & -1.10 & -1.11 & -1.08 \\
CH$_4$-Ne & DI & -0.07 & -0.14 & -0.07 & -0.06 & -0.10 & -0.08 & -0.09 \\
Ne-Ar & DI & -0.04 & -0.08 & -0.04 & -0.04 & -0.07 & -0.04 & -0.06 \\
        &&         &   &      &   &      &         & \\
MAE & & 0.12 &  0.21 & 0.06 &  0.11 &  0.09 &  0.05 &  \\	
MARE   &       & 16.2\%& 30.7\% & 11.3\% & 19.1\%&14.6\% & 9.1\% & \\
\hline
\multicolumn{9}{c}{$R=1.5R_0$}\\
HCN-ClF & CT  & -0.34 & -0.25 & -0.26 & -0.05 & -0.22 & -0.17 & -0.11 \\
NH$_3$-F$_2$ & CT & -0.19 & -0.21 & 0.17 & -0.16 & -0.21 & -0.18 & -0.21 \\
LiH-HF & DH & -0.71 & -1.04 & -0.64 & -0.51 & -0.72 & -0.60 & -0.65 \\
NH$_3$-H$_2$O & HB & -0.28 & -0.31 & -0.26 & -0.24 & -0.26 & -0.28 & -0.27 \\
(NH$_3$)$_2$ & HB & -0.21 & -0.21 & -0.20 & -0.19 & -0.21 & -0.21 & -0.21 \\
HCl-H$_2$S & DD & -0.35 & -0.37 & -0.30 & -0.23 & -0.31 & -0.29 & -0.31 \\
CH$_4$-Ne & DI & -0.01 & -0.02 & -0.01 & -0.01 & -0.02 & -0.01 & -0.02 \\
Ne-Ar & DI & -0.01 & -0.01 & -0.01 & -0.02 & -0.01 & -0.01 & -0.02 \\
        &&         &   &      &  &       &         & \\
MAE & & 0.04 &  0.08 & 0.03 & 0.04 &  0.03 &  0.03 & \\
MARE   &       & 37.5\%  & 33.6\% & 28.0\% & 22.9\% & 16.4\% & 17.9\% & \\
\end{tabular}
\end{ruledtabular}
\end{center}
\end{table*}

Inspection of the table and the figure shows that both the M1 and the M2
methods perform very well for the test set, being competitive with
CCSD(T)-23 and MP2.5/CBS, at all distances, and superior to MP2-23
and SCS(MI)-MP2/CBS.
In more detail, the M1 method seems to perform slightly better than M2 for hydrogen-bond
complexes at equilibrium distance, while M2 gives better results for
dispersion complexes and in general at displaced geometries. These differences
are however rather small and shall be considered with caution due to the
small dimension of the test set.
We recall that both methods have a computational cost
comparable with CCSD(T)-23 when a single bond distance (e.g. $R=R_0$)
is considered and smaller than it when several bond-distances are
involved. In particular the M2 method as a computational cost
almost comparable to MP2-23 (and SCS(MI)-MP2/CBS), 
but an accuracy close to CCSD(T)-23.
Note however that some the methods considered here
are based on extrapolation form triple-zeta quality basis
sets at most. Thus, it may happen that they display
shortcoming for some systems where this level of
basis set is not fully sufficient to describe the correlation
effects. This appears to be the case, for example of
the HCN-ClF complex which shows 
a quite larger relative error than
other complexes, especially at $R=1.5R_0$, possibly because of the
limitations of the aug-cc-pVTZ basis set to describe
correlation in the ClF molecule.

\begin{figure}
\includegraphics[width=\columnwidth]{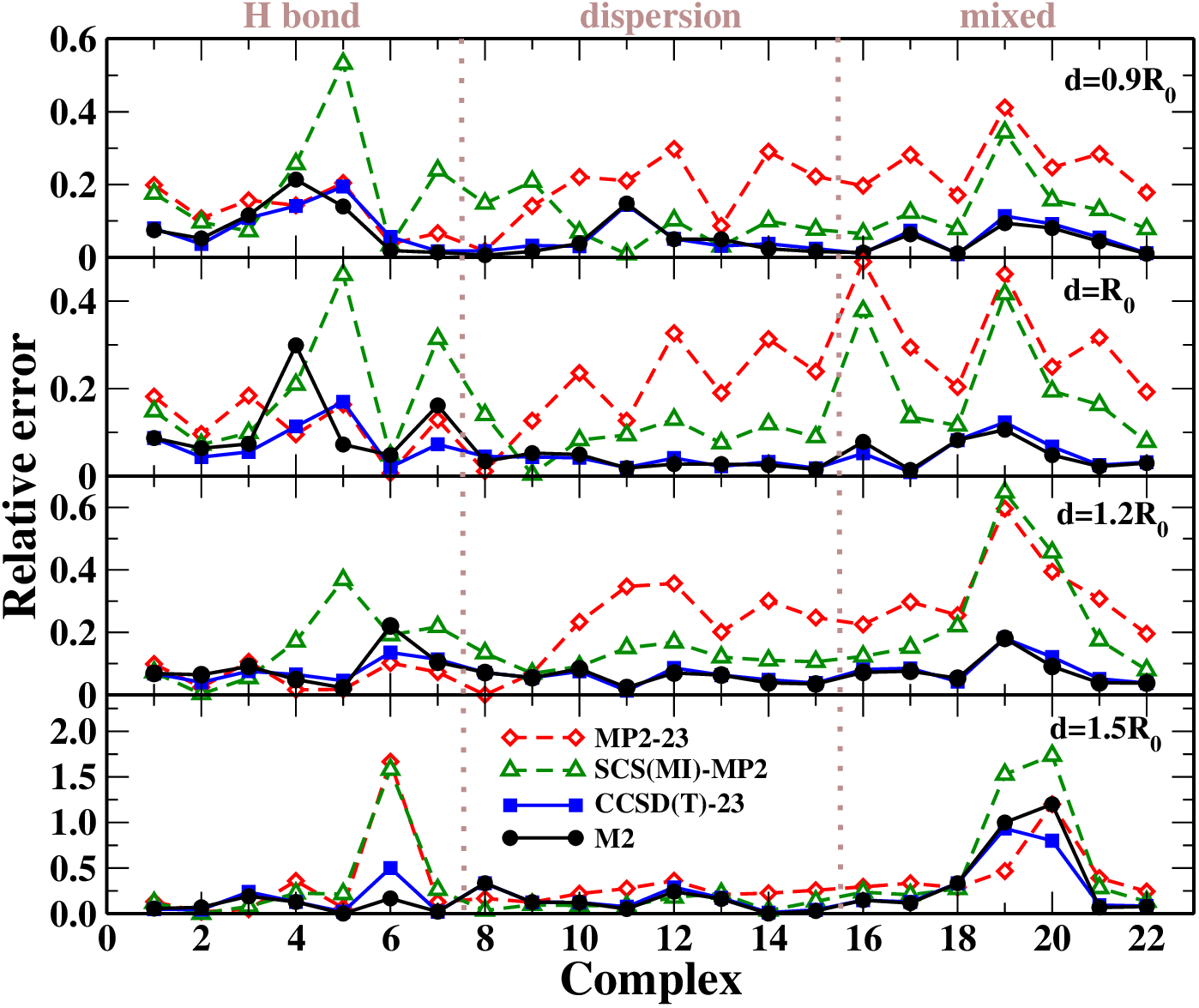}
\caption{\label{fig2}Absolute relative errors with respect to reference values \cite{grafova10} 
for the correlation interaction energies of non-covalent complexes of the S22x5 test set at different bonding distances $d$. The numbering of the complexes is: (1) 2-pyridoxine-2-aminopyridine, (2) adenine-thymine [WC], (3) ammonia dimer, (4) formamide dimer, (5) formic acid dimer, (6) water dimer, (7) uracil dimer [hb], (8) methane dimer, (9) ethene dimer, (10) benzene-methane, (11) benzene dimer [stack], (12) pyrazine dimer, (13) uracil dimer, (14) indole-benzene [stack], (15) adenine-thymine [stack], (16) benzene-ammonia, (17) benzene dimer [T shape], (18) benzene-water, (19) benzene-HCN, (20) ethene-ethyne, (21) indole-benzene [T shape], (22) phenol dimer. Note that complexes 6, 19, and 20 have interaction correlation energies close or smaller than 0.1 kcal/mol for $d=1.5R_0$, thus their relative errors are quite large in this case.}
\end{figure}

\begin{figure}
\includegraphics[width=\columnwidth]{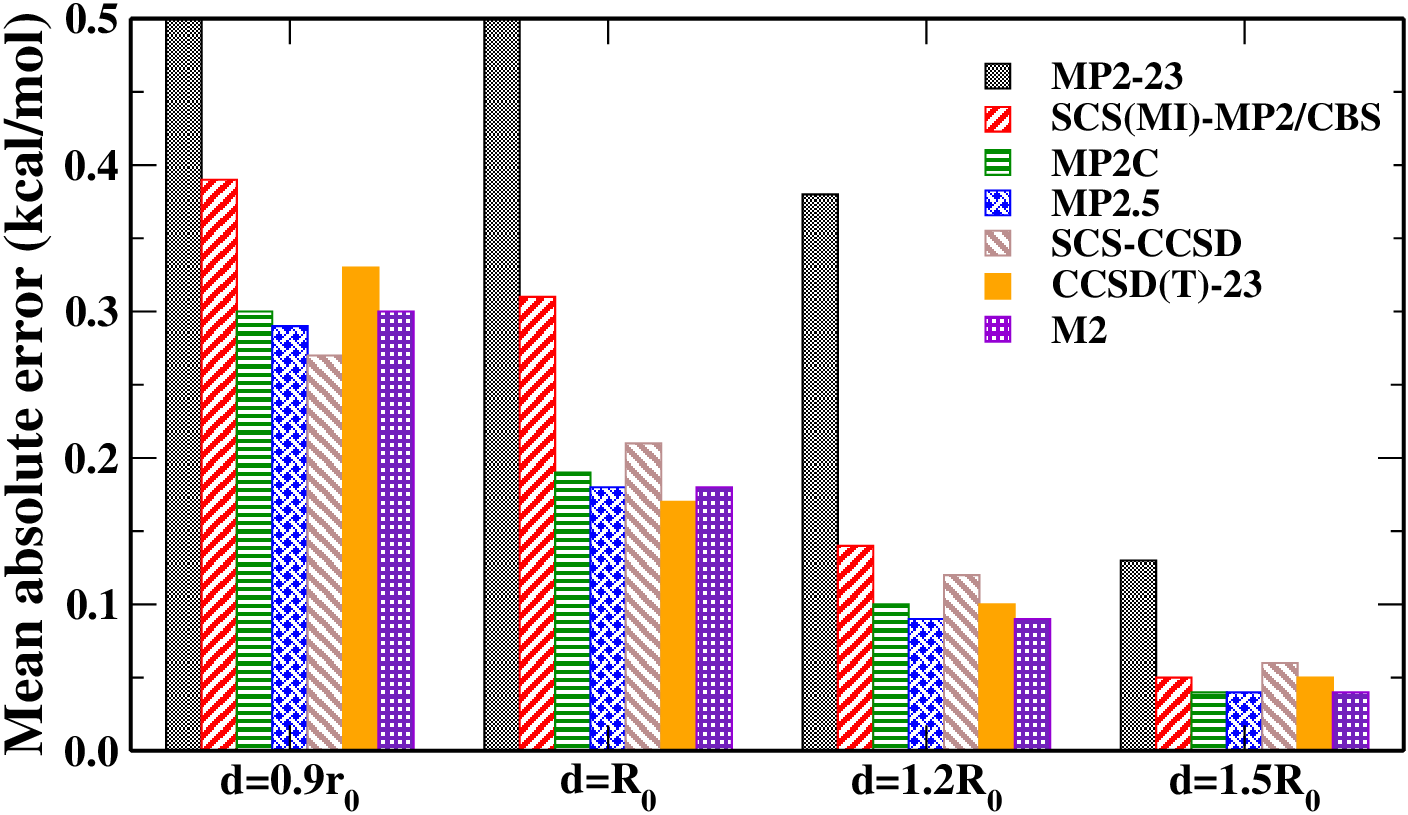}
\caption{\label{fig3} Mean absolute errors (kcal/mol) with respect to accurate values \cite{grafova10} for the total interaction energies of non-covalent complexes of the S22x5 test set at different bonding distances $d$ as obtained from several theoretical approaches. Accurate MP2C, MP2.5, and SCS-CCSD results are taken from Ref. \cite{grafova10}.}
\end{figure}

As a further and more extended test we consider
in Fig. \ref{fig2} (see also supporting information),
for the methods requiring at most triple-zeta
basis set calculations,
the absolute errors with respect to accurate values \cite{grafova10}
of correlation interaction energies
for the complexes of the S22 test set 
for $R_0$, 1.2$R_0$, and $1.5R_0$.
Accurate reference correlation interaction energies have been 
obtained by subtracting CBS Hartree-Fock values from the 
benchmark energies of Ref. \cite{grafova10}.
Statistics for these data are summarized in Table \ref{tab4}.
Finally, a comparison of total interaction energy errors with 
several literature results \cite{grafova10,mp2.5,pitonak10}
is also reported in Fig. \ref{fig3}.
\begin{table*}
\begin{center}
\caption{\label{tab4}Overall statistics, including mean absolute errror (MAE) in kcal/mol and mean absolute relative error, for the performance of different methods on the S22 test set at various equilibrium distances. The label SCS(MI) stands for SCS(MI)-MP2/CBS. The best result for each line is highlighted in bold style.}
\begin{ruledtabular}
\begin{tabular}{lllrrrr}
$R/R_0$ & Interaction & & MP2-23 & SCS(MI) & CCSD(T)-23 & M2 \\
\hline
0.9 & H-bond     & MAE  & 2.02 & 0.79 & 0.63 & \textbf{0.62} \\
    &            & MARE & 13.1\% & 20.1\% & \textbf{9.0\%} & \textbf{9.0\%} \\
    & Dispersion & MAE  & 0.71 & 0.81 & 0.31 & \textbf{0.30} \\
    &            & MARE & 18.6\% & 9.3\% & 4.6\% & \textbf{4.3\%} \\
    & Mixed      & MAE  & 2.25 & 1.04 & 0.39 & \textbf{0.37} \\
    &            & MARE & 25.3\% & 13.9\% & 5.2\% & \textbf{4.5\%} \\
    & All        & MAE  & 1.61 & 0.88 & 0.44 & \textbf{0.42} \\
    &            & MARE & 19.0\% & 14.2\% & 6.2\% & \textbf{5.9\%} \\
& & & & & & \\
1.0 & H-bond     & MAE  & 1.29 & 0.72 & \textbf{0.21} & 0.24 \\
    &            & MARE & 12.2\% & 19.3\% & \textbf{8.0\%} & 11.5\% \\
    & Dispersion & MAE  & 0.45 & 0.50 & \textbf{0.22} & 0.27 \\
    &            & MARE & 19.6\% & 9.2\% & 3.3\% & \textbf{3.1\%} \\
    & Mixed      & MAE  & 1.72 & 0.88 & \textbf{0.24} & 0.26 \\
    &            & MARE & 31.5\% & 21.1\% & 5.6\% & \textbf{5.4\%} \\
    & All        & MAE  & 1.12 & 0.69 & \textbf{0.22} & 0.26 \\
    &            & MARE & 21.1\% & 16.2\% & \textbf{5.5\%} & 6.5\% \\
& & & & & & \\
1.2 & H-bond     & MAE  & 0.56 & 0.27 & \textbf{0.14} & 0.15 \\
    &            & MARE & 6.2\% & 15.4\% & \textbf{7.8\%} & 8.9\% \\
    & Dispersion & MAE  & 0.17 & 0.28 & \textbf{0.08} & \textbf{0.08} \\
    &            & MARE & 21.9\% & 11.9\% & 5.6\% & \textbf{5.5\%} \\
    & Mixed      & MAE  & 0.73 & 0.41 & 0.18 & \textbf{0.16} \\
    &            & MARE & 32.4\% & 26.5\% & 8.6\% & \textbf{7.8\%} \\
    & All        & MAE  & 0.47 & 0.31 & \textbf{0.13} & \textbf{0.13} \\
    &            & MARE & 20.3\% & 17.6\% & \textbf{7.2\%} & 7.3\% \\
& & & & & & \\
1.5 & H-bond     & MAE  & 0.18 & 0.10 & \textbf{0.07} & \textbf{0.07} \\
    &            & MARE & 34.5\% & 35.4\% & 14.3\% & \textbf{9.0\%} \\
    & Dispersion & MAE  & 0.07 & 0.09 & \textbf{0.05} & \textbf{0.05} \\
    &            & MARE & 23.0\% & 10.9\% & 14.6\% & \textbf{13.3\%} \\
    & Mixed      & MAE  & 0.25 & 0.15 & 0.09 & \textbf{0.08} \\
    &            & MARE & 45.9\% & 62.7\% & \textbf{36.0\%} & 42.0\% \\ 
    & All        & MAE  & 0.16 & 0.11 & 0.07 & \textbf{0.06} \\
    &            & MARE & 33.9\% & 35.2\% & 21.3\% & \textbf{21.1\%}\\
\end{tabular}
\end{ruledtabular}
\end{center}
\end{table*}
The plots reported in the figure as well as 
the data of Table \ref{tab4}  confirm 
the results of Tables \ref{tab2} and
\ref{tab3}, showing that the M2 method
is competitive with the CCSD(T)-23 approach,
as well as with other high level methods 
(e.g. accurate MP2.5, MP2C, or SCS-CCSD) and can
clearly improve over the MP2-23 and SCS(MI)-MP2/CBS results.
We remark that this performance is obtained for all
distances by requiring only a single CCSD(T)/aug-cc-pVTZ calculation.
Moreover, although no clear systematic trend can be obtained for the
accuracy of different methods versus the bonding character type,
we remark that the observed behavior of M2 is rather consistent
for all the complexes and distances with CCSD(T)-23, being
even slightly superior for dispersion complexes. On the contrary,
MP2-23 appears rather accurate only for hydrogen bond complexes.
Finally, we remark that for $d=R_0$, where more
accurate reference data \cite{marshall11}
than those in Ref. \cite{grafova10} are available, the results
do not change substantially when the best reference values are used.
In fact, in this latter case the MAEs (MAREs) of
MP2-23, SCS(MI)-MP2/CBS, CCSD(T)-23, and M2 are
1.13 (19.5\%), 0.70 (14.5\%), 0.28 (6.0\%), and 0.31 (6.7\%),
respectively, in good agreement with the values reported in Table
\ref{tab4}.

\section{Conclusions}
We have proposed efficient scaling procedures to
compute accurate correlation interaction energies of
non-covalent complexes. Our methods are based
on the observed proportionality between MP2 and
CCSD(T) interaction energies as well as on
simple basis set extrapolation formulas.
In this way correlation interaction energies
of CCSD(T)/CBS quality can efficiently be obtained from
few MP2 and CCSD(T) calculations using small basis sets.
If several bonding distances are of interest,
as in PES scanning studies, even a lower computational
cost is achieved by exploiting the (almost) constancy
of the scaling factors with respect to
inter-molecular distance.
Thus, the present methods, and in particular the M2 one,
represent promising tools for future studies 
of large non-covalent systems, e.g. in
biochemistry.

Nevertheless, future developments can be foreseen starting from
the present results, especially to improve further
the overall computational efficiency. In this sense, it will be
particularly interesting to consider the
extension of the present methodology to 
scaled-opposite-spin (SOS) MP2 calculations \cite{sosmp2}.
In fact, the SOS-MP2 method can be implemented with a favorable
$\mathcal{O}(N^4)$ scaling (as opposed to the 
$\mathcal{O}(N^5)$ of conventional MP2) and also
displays a proportionality with 
accurate correlation results \cite{cos_scal,sosmp2,oepsos1,oepsos2}.

\section*{Acknowledgments}This work was partially supported by the National
Science Center under Grant No. DEC-2013/11/B/ST4/00771.

\appendix

\section{Rationalization of Eq. (\ref{e3})}
\label{appa}
Following Refs. \cite{truhlar1,truhlar2} we can write
\begin{eqnarray}
\mathcal{E}_\mathrm{CCSD(T)}^{(n)} & = & \mathcal{E}_\mathrm{CCSD(T)}^{(\infty)} +Bn^{-\beta} \\
\mathcal{E}_\mathrm{MP2}^{(n)} & = & \mathcal{E}_\mathrm{MP2}^{(\infty)} +Cn^{-\gamma}\ ,
\end{eqnarray}
where $B$, $C$, $\beta$, and $\gamma$ are constants.
Taking the ratio of the two equations we find
\begin{eqnarray}
\nonumber
c^{(n)} & = & \frac{\mathcal{E}_\mathrm{CCSD(T)}^{(n)}}{\mathcal{E}_\mathrm{MP2}^{(n)}} = \frac{\mathcal{E}_\mathrm{CCSD(T)}^{(\infty)} +Bn^{-\beta}}{\mathcal{E}_\mathrm{MP2}^{(\infty)} +Cn^{-\gamma}} = \\
& = & \frac{\mathcal{E}_\mathrm{CCSD(T)}^{(\infty)}/\mathcal{E}_\mathrm{MP2}^{(\infty)} +\tilde{B}n^{-\beta}}{1 +\tilde{C}n^{-\gamma}}\ ,
\end{eqnarray}
with $\tilde{B}=B/\mathcal{E}_\mathrm{MP2}^{(\infty)}$ and $\tilde{C}=C/\mathcal{E}_\mathrm{MP2}^{(\infty)}$ (note these are constants
with respect to $n$).
Now observing that
\begin{equation}
\left|\tilde{C}n^{-\gamma}\right|\approx\left|\frac{\mathcal{E}^{(n)}_\mathrm{MP2}}{\mathcal{E}_\mathrm{MP2}^{(\infty)}}-1\right|< 1\ ,
\end{equation}
we can write
\begin{eqnarray}
\nonumber
c^{(n)} & \approx & c^{(\infty)} + \tilde{B}n^{-\beta}\left[1-\tilde{C}n^{-\gamma} + \mathcal{O}\left(\tilde{C}^2n^{-2\gamma}\right)\right] \approx\\
& \approx & c^{(\infty)} + \tilde{B}n^{-\beta} - c^{(\infty)}\tilde{C}n^{-\gamma}\ .
\end{eqnarray}
Finally, considering that $\beta\sim\gamma$ (see  
Refs. \cite{truhlar1,truhlar2}) we can define
two new constants
$\beta\sim\gamma\sim\alpha$ and $
A=\tilde{B}-c^{(\infty)}\tilde{C}$ and obtain
\begin{equation}
c^{(n)} \approx c^{(\infty)} + An^{-\alpha}\ .
\end{equation}

\end{document}